# Towards Accurate Performance Modeling of RISC-V Designs

Odysseas Chatzopoulos     George-Marios Fragkoulis     George Papadimitriou     Dimitris Gizopoulos

Department of Informatics and Telecommunications
University of Athens, Greece
{od.chatzopoulos | gm.fragkoulis | georgepap | dgizop}@di.uoa.gr

## ABSTRACT

Microprocessor design, debug, and validation research and development are increasingly based on modeling and simulation at different abstraction layers. Microarchitecture-level simulators have become the most commonly used tools for performance evaluation, due to their high simulation throughput, compared to lower levels of abstraction, but usually come at the cost of loss of hardware accuracy. As a result, the implementation, speed, and accuracy of microarchitectural simulators are becoming more and more crucial for researchers and microprocessor architects. One of the most critical aspects of a microarchitectural simulator is its ability to accurately express design standards as various aspects of the microarchitecture change during design refinement. On the other hand, modern microprocessor models rely on dedicated hardware implementations, making the design space exploration a time-consuming process that can be performed using a variety of methods, ranging from high-level models to hardware prototyping. Therefore, the tradeoff between simulation speed and accuracy, can be significantly varied, and an application's performance measurements uncertain.

In this paper, we present a microarchitecture-level simulation modeling study, which enables as accurate as possible performance modeling of a RISC-V out-of-order superscalar microprocessor core. By diligently adjusting several important microarchitectural parameters of the widely used gem5 simulator, we investigate the challenges of accurate performance modeling on microarchitecture-level simulation compared to accuracy and low simulation throughput of RTL simulation of the target design. Further, we demonstrate the main sources of errors that prevent high accuracy levels of the microarchitecture-level modeling.

## 1 INTRODUCTION

Cycle-accurate (performance) simulators are widely used in the early stages of microprocessor design, as well as in several research studies, e.g., for performance evaluation, reliability assessment, etc. [1] [2] [3] [4]. Microprocessor model validation and accuracy is among the most common and significant problems that designers and researchers are regularly dealing with. Accurate simulation infrastructures are essential for design space exploration to ensure that the most efficient solution will be chosen. Since exact technology libraries for new microprocessor architectures are not available at early design stages, microprocessor architects (and researchers) usually simulate their designs using models at different abstraction layers, ranging from register-transfer level (RTL) to full-system level (microarchitecture level). An RTL model can simulate the real hardware design in a cycle-accurate manner, making it much more precise than higher-level abstractions, such as microarchitecture-level simulations. RTL simulation also allows accurate measurements for hardware area, and power consumption using commercial Computer-Aided Design (CAD) tools without introducing modeling errors [5]. However, RTL simulation of a complete microprocessor model using real-world applications or benchmarks, adds significant delays on the workflow, primarily due to very low simulation throughput of RTL. For example, in [6] the authors show that the magnitude of difference of execution time of RTL simulation can be as large as 7x more than the microarchitecture-level simulation time. Figure 1 summarizes an abstract view of the relative simulation speed and accuracy among the most common simulation methods at different abstraction layers.

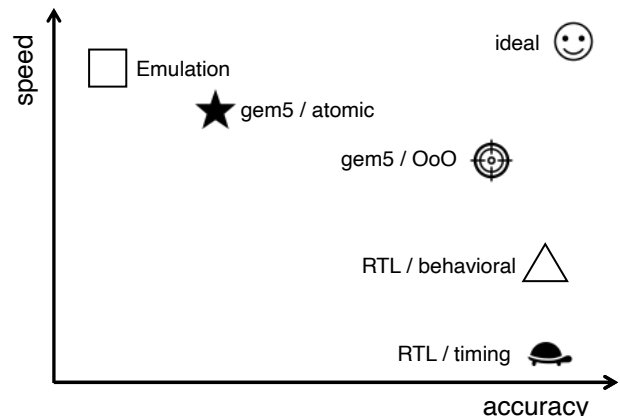

**Figure 1: Simulation speed vs. accuracy for the widely-used methods on different abstraction layers and simulation modes.**

Due to high simulation throughput, microarchitecture-level simulators (such as the widely used gem5 simulator [1]) are heavily employed [6] [7] [8], mainly for performance evaluation studies at the cost of accuracy loss [5]. Microarchitecture-level simulators model designs of different architectures at a higher abstraction layer than RTL. Consequently, they can evaluate real-world applications in reasonable simulation time, however, they fail to model accurately the real hardware providing design inconsistencies, and thus, clock cycle inaccuracies. Black and Shen in [7] categorize the sources of simulation errors into three distinct groups: (i) *modeling errors*, when the simulator developers incorrectly model the desired functionality; (ii) *specification errors*, when the developers are unaware of the functionality being modeled or must speculate about it; (iii) *abstraction errors*, when the developers abstract or fail to incorporate some details of the modeled design. Modelling errors can usually be corrected when the simulator is gradually improved through its continuous maintenance. On the other hand, specification and abstraction errors tend to be more persistent and difficult to correct, since certain aspects of the system may be unavailable to the public, or abstractions may simplify simulator code and speed up simulations [5]. However, in case of RISC-V design implementations, specification errors can be also overcome (along with the modeling errors), since there are several RISC-V open-source designs.

As shown in Figure 1, there is no ideal simulation method which can bridge the gap between simulation speed and accuracy, although there have been substantial research efforts trying to overcome this challenge, usually by compromising one of these two aspects. For example, Graphite multicore simulator [8] aims to improve the simulation speed by compromising the simulation accuracy. Sniper [9] simulator speeds up simulation by using high level models of cores instead of performing cycle accurate simulation. Zsim simulator [10] reduces the time for detailed simulation by employing dynamic binary translation for instruction driven timing models. All these simulators utilize system-level description of the design and lack the accuracy of RTL. Along the same lines, ExtraTime [11] is a complex simulation platform for simulating and modeling power and area at the microarchitectural level to investigate the effect of aging. It is also based on limited accuracy for critical performance parameters, because it lacks the hardware-level accuracy required for performance characterization provided by the RTL.

In this paper, we present a microarchitecture-level simulation modeling approach based on gem5 simulator, which enables as accurate as possible performance modeling of a RISC-V out-of-order (OoO) superscalar microprocessor core while retaining the high throughput of microarchitecture-level simulation. We demonstrate the challenges for adjusting several microarchitectural parameters of the microarchitecture model based on the widely used gem5 simulator. Considering the RSD microprocessor core as our baseline RTL model [12], we show which are the main sources of error in performance modeling through microarchitecture-level simulation compared to the RTL simulation of the target design. RSD is a compact open-source RISC-V OoO superscalar microprocessor core, which can be synthesized for FPGAs. Further, we demonstrate the benefits and high accuracy levels that can be achieved, by avoiding the low simulation throughput of RTL simulation.

The main contributions of this study are: (1) the detailed discussion about the challenges and sources of error between microarchitecture-level and RTL simulations, and (2) the modeling validation of two ISAs with subtle differences, i.e., the 32-bit and 64-bit RISC-V ISAs. The purpose of this study is to present some preliminary evaluation results of such an important problem of the validation of microarchitectural simulators, which are the main vehicle for the most of the research studied in the field of computer architecture.

## 2 BACKGROUND & RELATED WORK

In the early designing phases, new microprocessor designs are typically evaluated using RTL or microarchitecture-level simulators. Although microarchitecture-level simulators employ fast high-level models (i.e., high simulation throughput), they usually provide less accuracy compared to RTL models. On the contrary, RTL simulators provide cycle-accurate modeling of the underlying hardware, but due to their very low simulation throughput, it is infeasible to simulate long and realistic workloads. In this section, we briefly describe the background and motivation of this study and the most recent related works.

### 2.1 Background and Motivation

Computer architects and researchers have widely used cycle-accurate microarchitectural simulation for several kinds of evaluation purposes, such as performance evaluation. Microarchitecture-level simulators are modular and simple to use, while RTL implementation is difficult to change. As a result, high-level software simulation is still a valuable tool for guiding system design in the early stages, before RTL design begins. However, for two major reasons, microarchitecture-level simulation has been a bottleneck in several studies. The first reason is that microarchitecture-level simulators should be thoroughly tested against RTL designs and real-world systems. This is only possible if new prototypes share many similarities with the current simulated hardware or similar designs from previously tested design cycles. Microarchitecture-level simulators should be fine-tuned whenever any modifications to RTL designs are made. Furthermore, microarchitectural simulators should be thoroughly tested against silicon implementations that run real-world software. Otherwise, the abstraction and modeling of the target systems may introduce various types of evaluation errors [5]. Hardware design trends developing heterogeneous or other complex System on Chips (SoCs) usually using custom hardware accelerators have made microarchitecture-level simulator validation more challenging as it has become more difficult to find an existing device to validate the simulator against. On the other hand, RTL simulation provides very low simulation throughput of realistic workloads with very complicated hardware designs, when complete executions of programs are mandatory. The low simulation throughput of RTL is becoming even worse when running full-system simulations.

There are usually two fundamental approaches that are commonly used in an effort to overcome the previously defined challenges. First, simulation sampling is a widely-known and widely-used method of reducing the simulation time [13]. Through an extensive phase analysis on dynamic basic-block traces, simulation sampling provides several strictly-defined simulation points (usually referred to as simpoints). The underlying assumption is that dynamic software execution is made up of short periods of phases, with each phase exhibiting similar microarchitectural activity, and therefore similar instructions per cycle (IPC), when repeated. This approach generates simulation points that are based around the most commonly used basic blocks. Since these phase periods can be long enough and the performance characteristics of each phase can be dependent on the system's dynamic state, this assumption cannot provide any guarantees that the outcomes will be correct for every single case.

The other fundamental approach that aims to significantly improve the simulation time is the FPGA-accelerated simulation of an RTL design [14] [15]. The FPGA-based performance simulators are orders of magnitude faster than microarchitecture-level or RTL simulators, because execution takes place on a real silicon chip; however, they require manually description of abstract models in RTL, which may be more difficult than writing the source RTL design. Another major challenge is the mapping of simulation models into the FPGA platform, including timing and software models and the transformation from RTL model. Since any of these approaches have several important limitations, in this work, we study the microarchitecture-level simulation, which can execute long and realistic workloads with very low implementation and modeling effort, towards an accurate performance modeling, specifically for RISC-V designs.

## 2.2 Related Work

For performance characterization and design modeling, gem5 [1] is a commonly used microarchitecture-level simulator. In [16], gem5 is used to simulate in-order and out-of-order Arm microprocessors. In [17], gem5 is extended to support VLIW instruction, and the modeling infrastructure is validated using an RTL simulator as a reference. Zamn *et al.* in [6] propose a cross-layer approach using the gem5, which enables accurate power estimation by integrating components from system-level and RTL simulation of the target design. Butko *et al.* in [1] evaluate the accuracy modeling of gem5 and show the inaccuracy levels compared to a dual-core ARM Cortex-A9 real microprocessor device.

Similar to these studies, in the context of RISC-V designs, Roelke and Stan in [18] implemented the RISC-V ISA in gem5 and validate its performance statistics against the Chisel simulation and FPGA. Ta *et al.* in [19] present functional and timing validation of multicore RISC-V designs on gem5. Kim *et al.* in [14] and [15] present an FPGA-accelerated methodology for simulation-based RTL verification of RISC-V designs and an evaluation methodology using RTL designs running real-world workloads in FPGA simulation to evaluate performance, power and energy. There are also several studies that present the sources of errors in microarchitecture-level simulators. Gutierrez *et al.* in [5] validate gem5 simulator and quantify the error magnitude against a real hardware platform. Brooks *et al.* in [2] investigate the primary sources of error of microarchitecture-level simulation and present how design tradeoff studies can accommodate some inaccuracy since relative inaccuracy has no impact on the target design. Desikan *et al.* in [20] verify the accuracy of a high-level timing simulator against actual hardware.

## 3 MODELING INFRASTRUCTURE

In this section, we provide an overview about the RSD microprocessor core, which is used as the reference RTL model in this study, and also discuss about the most important features of the gem5 microarchitecture-level simulator.

## 3.1 RSD: A Reference Model

RSD is an open-source RISC-V OoO microprocessor core optimized for FPGA, which provides high performance by supporting advanced microarchitectural features such as speculative OoO load and store execution, a memory dependence predictor, speculative scheduling, and a non-blocking cache. RSD improves resource efficiency by minimizing the used FPGA resources, by supporting multiport RAM arrays for several RAM-based components, like Reorder Buffer (ROB), Physical Register File, etc. [12]. The RSD pipeline is structured using three basic blocks: (i) the front-end block, (ii) the scheduling block, and (iii) the execution block. The front-end block of RSD microarchitecture fetches and decodes instructions, in an in-order manner according to the program's instruction order, from the L1 instruction cache, and supports the *gshare* branch predictor. For instructions sent from the front-end block, the scheduling block provides instruction-level parallelism (ILP), and issues instructions to the execution block in an out-of-order manner. The rename unit, dispatch unit, issue queue, and reorder buffer are the main components of the scheduling block. Instructions sent from the scheduling block are executed by the execution block. The execution block consists of a physical register file (of 64 registers count) and a load/store unit, which contribute to the execution and speculative execution of memory instructions out of program order. Table 1 shows the most important microarchitectural parameters of the RSD core.

Compared to the state-of-the-art RISC-V microprocessors models, like Rocket [21], RSD supports a more aggressive design since Rocket is an in-order, single-issue scalar microprocessor that includes a six-stage integer pipeline. BOOM [22] is another state-of-the-art RISC-V microprocessor, which is a high-performance out-of-order superscalar microprocessor core, providing several complex features which make RTL simulations even slower than of Rocket's and RSD's. BOOM supports a unified physical register file with configurable fetch widths, issue widths, and instruction window sizes. It supports full branch speculation using a branch target buffer, and a parameterizable backing predictor.

## 3.2 Gem5 Microarchitectural Simulator

Gem5 is a widely used open-source microarchitecture-level simulator, which supports a variety of instruction set architectures (ISA), microprocessor and memory system models. Each microprocessor model corresponds to a different abstraction layer, and thus, different level of simulation throughput and accuracy. For example, the atomic model (as shown in Figure 1) comes with low accuracy but high throughput, while the detailed out-of-order model (OoO) provides lower throughput but accurate simulations. System-call emulation (SE) and full-system (FS) simulation modes are available for gem5 microprocessor models. During the simulation in SE mode (which is the one used in this work), gem5 does not load an operating system, and system calls are emulated by the host system. The FS mode, on the other hand, executes both user-level and kernel-level instructions when simulating a full system by loading an operating system into the simulator. The operating system is responsible to simulate all system calls and to perform virtual-to-physical address translations.

| Parameter | Value |
|---|---|
| Pipeline | OoO |
| L1 data / instruction cache | 4 KB / 4 KB (2-way) |
| Cache line size | 8 Bytes |
| Replacement policy | Tree-PLRU |
| L1 hit latency | 1 clock cycle |
| L1 miss latency | 100 clock cycles |
| Fetch/Decode/Rename width | 2 |
| Issue width | 5 |
| Writeback width | 5 |
| Commit width | 2 |
| Reorder Buffer | 64 |
| MSHR entries | 2 |
| Branch predictor | gshare (2048 History Table) |
| Branch Target Buffer entries | 1024 |
| Load/Store Queues entries | 16 |
| Physical Register File | 64 registers |

**Table 1: Microarchitectural parameters of the RSD reference model.**

### 3.3 RISC-V ISAs Compatibility

The base instruction set architectures of RISC-V are RV32I and RV64I. RSD implements RV32I, while gem5 implements RV64I. The actual difference between these two ISAs is the register width; RV32I has 32bit registers, while RV64I has 64bit registers. Both ISAs support the same instruction format and width, i.e., all instructions are 32-bit and the opcodes are the same between these two ISAs. The main difference of the instruction format between the two ISAs is the different width of the data (32-bit vs. 64-bit). Larger data widths of the RV64I ISA compared to RV32I adds a few more instructions on the instruction set to deal with the larger data sizes. For example, assume the add instruction, in which the same opcode and instruction format is using for both RV32I and RV64I, but the former has to deal with 32-bit data sizes, while the latter with 64-bit data sizes. Further, RV64I supports more instructions (i.e., extra opcodes) to cope with 32-bit data sizes (e.g., the *addw* instruction). In general, RV64I supports additional instruction variants for manipulating 32-bit values, indicated by a 'W' suffix to the opcode. The same exists also for load and store instruction, which provide the same opcodes and instruction format for both RV32I and RV64I, but they have to deal with 32bit and 64-bit values, respectively [23].

Since both 32-bit and 64-bit ISAs share only a few minor differences according to the previous discussion, we challenge ourselves to validate the microarchitecture-level accuracy of a 64-bit ISA compared to an RTL which implements the 32-bit ISA of RISC-V. These two RISC-V ISAs have subtle differences, specifically, they differ only to the data size manipulation. To avoid the potential inaccuracies in performance measurements of our study, we use programs which use integer numbers of 32-bit size for both ISAs and microarchitecture models of this study. Further, for both ISAs we use the same version of the RISC-V cross-compiler, i.e., gcc v10.2 without any optimizations (-O0).

## 4 EXPERIMENTAL METHODOLOGY & RESULTS

In this section, we first discuss about the experimental methodology followed by this study and we present our evaluation results. To compare the microarchitecture-level simulation accuracy to the RTL reference model (i.e., the RSD in this study), we initially configured all the available microarchitectural parameters of the gem5 simulator considering the parameters of the RSD, as shown in Table 1. By doing so, we can compare the two identical models, and provide fundamental observations about the potential inaccuracies that can occur due to limited hardware details implemented on the gem5.

### 4.1 Benchmarks and Simulation Speed

In this study, we mainly use custom-developed benchmarks, which target to specific microarchitectural components. The reason is twofold: first, to study and evaluate the microarchitecture-level simulation accuracy, and second to retrieve some specific microarchitectural parameters of the RSD reference model that were not available. For example, for the verification of the integer pipeline, we developed the *IntegerStress* benchmark, which stresses the integer functional units only. Along with custom-developed benchmarks, we also use 3 benchmarks from MiBench suite (*qsort* and *stringsearch* with small and large input datasets) [24]. Table 2 presents the benchmarks used in this study along with a representative description which describes each benchmark.

By using these benchmarks, we first evaluate the simulation speed between the gem5 and RSD. Note that in gem5 we use the detailed out-of-order mode, while in RSD we rely on the fast and cycle-accurate behavioral simulation model, which is significantly faster than a detailed timing simulation. As shown in Figure 2, and specifically in the dotted line, depending on the workload and the application's characteristics, the speedup of microarchitecture-level simulation ranges between 5x and 20x compared to RTL behavioral simulation. Specifically, the fewer memory operations a program includes, the faster gem5 simulation is. As we discuss in the next subsection, the simulated memory system is one of the two main obstacles, which contribute to the highest levels of the microarchitecture-level simulation inaccuracy.

### 4.2 Runtime Accuracy

The gem5 modeling is as accurate as possible compared to the reference RTL model. More specifically, both integer and memory pipelines are completely matched. We employ Konata[1] to verify these features as well as the *IntegerStress* benchmark described in the previous subsection that uses only arithmetic instructions without any cache misses. Konata is an instruction pipeline visualizer, which allows us to pinpoint any divergencies of the pipeline flow between gem5 and RSD models. Konata is an essential tool which has significantly contributed to this study. Figure 3 presents the ratio of clock cycles (left diagram) and the ratio of instructions (right diagram), between gem5 and RSD. In the rightmost part of each diagram, we can see also the arithmetic, geometric, and harmonic means. As we can see, the ideal ratio would be equal to 1 (the red horizontal line). On the left diagram of Figure 3 we can see that *BubblesortC* and *StringSearchSmall* benchmarks show the highest difference in clock cycles among all benchmarks. In *BubblesortC*, gem5 reports 36% more clock cycles than the RTL model, while in *StringSearchSmall* gem5 reports 35% less clock cycles that the RTL model.

---
[1] https://github.com/shioyadan/Konata

| Benchmark | Description |
|---|---|
| Bubblesort C | Sort a 250-entry array with integer values using the Bubblesort algorithm. Assembly version of Bubblesort was written because C version shows considerable difference in committed instructions between the two ISAs |
| Bubblesort Asm | |
| MemoryRandom | Study the behavior of memory on random accesses |
| FibSlow | FibSlow runs for the 20th term of Fibonacci sequence. FibFast computes all terms in [1,45] and it reduces exponential complexity to linear |
| FibFast | |
| IntegerStress | Stresses the integer functional units |
| BranchMisNever | Always taken branch in a loop |
| BranchMisRandom | Specify branch misprediction latency studying gshare's behavior on branches using random conditions |
| StringSearchLarge | Searches for given words in phrases using a case insensitive comparison algorithm [24] |
| StringSearchSmall | |
| Qsort | Sorts a large array of strings into ascending order using the quick sort algorithm [24] |

**Table 2: Benchmarks used in this study along with a brief description.**

However, *IntegerStress*, *BranchMisRandom*, *MemoryRandom*, and *BranchMisNever*, show marginal difference in clock cycles (up to 6%), since they are all very close to the red horizontal line (i.e., the ideal ratio). Considering that the majority of performance evaluation studies employ a huge set of benchmarks and usually provide an average of the execution times of all benchmarks to be easy to compare several features, averaging can be assumed as an important factor for a fair comparison. As we can see in Figure 3, by averaging all benchmarks (the rightmost columns of the left graph), the clock cycle ratio is significantly close to the ideal ratio of 1. Note that this is true for any averaging method, such arithmetic, geometric or harmonic mean, which are the most commonly used methods in the majority of performance studies.

On the other hand, as we can see in the right diagram of Figure 3, the committed instruction ratio of both gem5 and RSD are virtually the same for the most cases. Perfectly matching the committed instructions is a real challenge because we noticed that major differences produced due to the different implementations of the *gshare* branch predictor. Therefore, we developed the *gshare* branch predictor in gem5 by directly translating the SystemVerilog implementation of RSD into C++ code. To this end, we developed the benchmarks targeted to the branch predictor (i.e., *BranchMisRandom* and *BranchMisNever*). *BranchMisRandom* specifies the branch misprediction latency in combination with *BranchMisNever* benchmark studying the behavior of branches with random conditions. By doing so, we were able to match the committed instructions between gem5 and RSD to the highest extent possible.

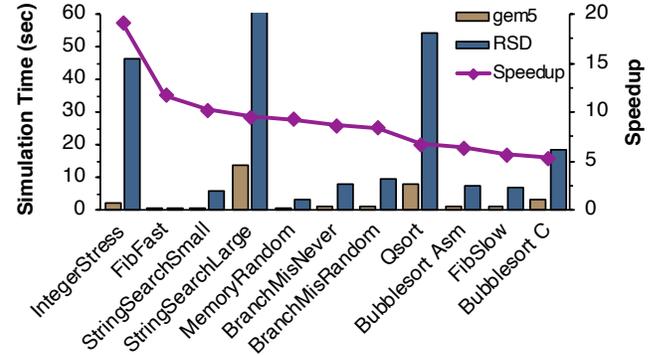

**Figure 2: End-to-end benchmark simulation time in seconds between gem5 and RTL simulation. The dotted line shows the speedup for each benchmark.**

### 4.3 Abstraction Errors & Error Magnitude

As we discussed in the previous subsection, although the committed instructions of gem5 are very close to RSD's, the clock cycles ratio can show some divergences, even if the two models (gem5 and RTL) are as close as possible. To provide observations about the differences in clock cycles between gem5 and RTL models, it is essential to study the most important performance characteristics of the microarchitecture model in a fine-grained manner. To this end, we studied several microarchitectural statistics, such as branch mispredictions, cache misses, MSHR hits, memory latency, etc., in an effort to correlate what is the possible factor (or factors) that affects the accuracy of microarchitectural modeling. In the majority of the studied statistics, we observed that both gem5 and RTL models report the same measurements or the same trends. However, a significant observation of our study is that the memory system and the branch predictor are the main factors which affect the accuracy of the microarchitecture-level modeling and the main sources of errors in microarchitecture-level performance modeling. Figure 4 presents the number of mispredictions per kilo instructions (MPKI) for each benchmark correlating to the clock cycles ratio (left diagram), and the memory accesses per kilo instructions (MAKI) for each benchmark correlating to the clock cycles ratio (right diagram).

As we can see in Figure 4, the four leftmost benchmarks (*BubblesortC*, *FibSlow*, *Qsort*, *BubblesortAsm*), in which microarchitecture-level modeling shows more clock cycles than the RTL modeling (i.e., there is a level of inaccuracy), provide the highest misprediction rate and the highest memory accesses (the yellow dotted line in the two graphs). Furthermore, the three rightmost benchmarks (*FibFast*, *StringSearchLarge*, *StringSearchSmall*), in which microarchitecture-level modeling shows less clock cycles than the RTL modeling (i.e., there is a level of inaccuracy), provide very low misprediction rate, but similarly to the leftmost benchmarks, very high memory accesses. On the other hand, benchmarks in the middle (*IntegerStress*, *BranchMisRandom*, *MemoryRandom*, *BranchMisNever*), which show the best accuracy of microarchitecture-level modeling, provide extremely low mispredictions and memory accesses rates, with *MemoryRandom* benchmark being an outlier. Led by these observations, we conclude that branch predictor and memory system are the most significant factors which can affect the accuracy levels of the microarchitecture-level modeling (these are both belong to the abstraction errors; meaning that the developers of the microarchitectural simulator abstract or fail to incorporate some details of the modeled design, as we discussed in Section 1). Both factors can contribute the most to the total performance of the microprocessors, and thus, implementation divergences between microarchitecture-level and RTL simulation can affect the accuracy of the modeling, and in many times, not in an obvious way. Therefore, it should be further investigated the reasons about how the branch predictor and memory system can be improved.

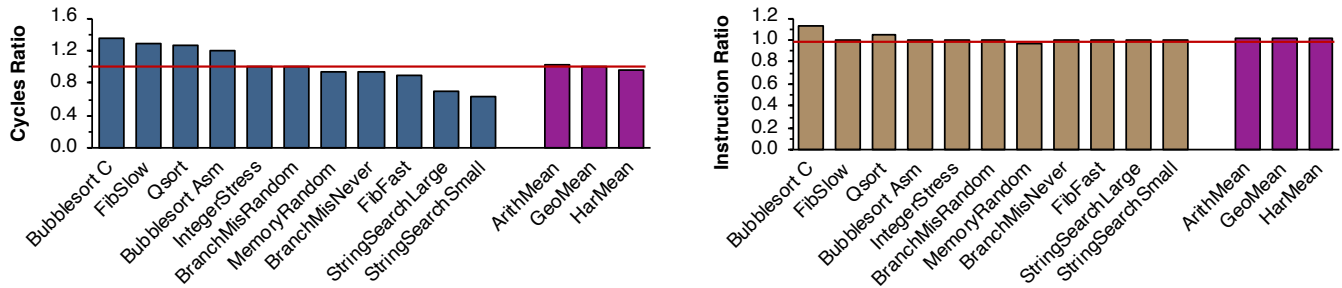

**Figure 3: Clock cycles ratio (left) and instruction ratio (right) between gem5 and RTL simulation for RSD model.**

## 5 CONCLUSION & FUTURE WORK

In this paper, we focus on the accuracy validation of performance modeling of microarchitecture-level simulation by employing a state-of-the-art RISC-V RTL design. We presented our methodology and the main characteristics of the gem5 simulator and the RSD design, which are both used as the main vehicle of this study. The scope of this work is to present the challenges of performance modeling using the microarchitecture-level simulation, in the context of accuracy. Through our experimental results, we show that although gem5 is accurately configured based on the microarchitectural characteristics of RSD, the clock cycles of some benchmarks, between gem5 and RSD, can be different up to 36% and 18% on average among all benchmarks due to the abstraction errors of microarchitecture-level modeling. However, the arithmetic, geometric and harmonic mean show that considering all benchmarks, the average of clock cycles of microarchitecture-level simulation can be as accurate as the RTL simulation. Further, we show that the main sources of error in the performance differences between gem5 and RSD seem to be the behavior of the branch prediction unit and the memory system. In future work, we primarily intend to investigate the way that these sources of errors affect the microarchitecture-level accuracy, and to provide solutions about the changes need to be performed to eliminate the differences compared to RTL simulation. Moreover, another future work direction will be the accuracy validation of performance modeling of microarchitecture-level simulation of an RTL design (e.g., BOOM/SonicBOOM [25]), using the full-system setup (i.e., with operating system support), and running longer and more representative workloads.

## ACKNOWLEDGMENTS

The authors are grateful to Prof. Ryota Shioya, from the University of Tokyo, for his insightful guidance and great support on the RSD implementation details.

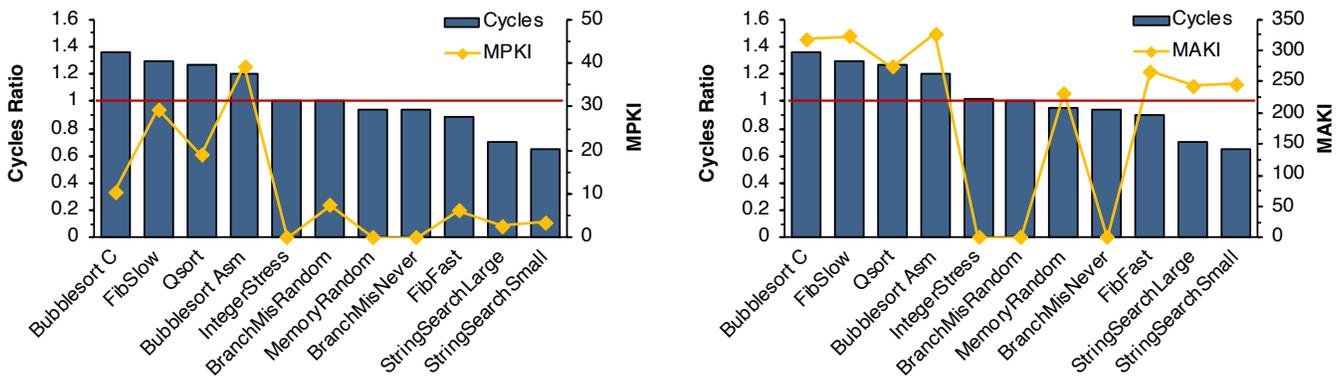

**Figure 4: Mispredictions per kilo instructions (MPKI) in correlation to the clock cycles ratio (left diagram) and memory accesses per kilo instructions in correlation to clock cycles ratio (right diagram).**